\documentclass[structabstract]{aa}  
\usepackage{graphicx}
\usepackage{txfonts}
\begin{document}
   \title{Analysis of new high-precision transit light curves of WASP-10~b: starspot occultations, small planetary radius, and high metallicity\thanks{Based on observations collected at the Centro Astron\'omico Hispano Alem\'an (CAHA), operated jointly by the Max-Planck Institut f\"ur Astronomie and the Instituto de Astrofisica de Andalucia (CSIC).}}
   \titlerunning{Analysis of new high-precision transit light curves of WASP-10~b}

   \author{G.~Maciejewski\inst{1,2},
           St.~Raetz\inst{2},
           N.Nettelmann\inst{3},
           M.~Seeliger\inst{2}, 
           Ch.~Adam\inst{2},
           G.~Nowak\inst{1}, and
           R.~Neuh\"auser\inst{2}
          }

   \institute{Toru\'n Centre for Astronomy, Nicolaus Copernicus University, 
              Gagarina 11, PL--87100 Toru\'n, Poland\\
              \email{gm@astri.uni.torun.pl}
         \and
              Astrophysikalisches Institut und Universit\"ats-Sternwarte, 
              Schillerg\"asschen 2--3, D--07745 Jena, Germany 
         \and
              Institut f\"ur Physik, Universit\"at Rostock, D--18051 Rostock, Germany
             }
   \authorrunning{G.~Maciejewski et al.}
   \date{Received ...; accepted ...}

 
  \abstract
   {The WASP-10 planetary system is intriguing because different values of radius have been reported for its transiting exoplanet. The host star exhibits activity in terms of photometric variability, which is caused by the rotational modulation of the spots. Moreover, a periodic modulation has been discovered in transit timing of WASP-10~b, which could be a sign of an additional body perturbing the orbital motion of the transiting planet.}
   {We attempt to refine the physical parameters of the system, in particular the planetary radius, which is crucial for studying the internal structure of the transiting planet. We also determine new mid-transit times to confirm or refute observed anomalies in transit timing.}
   {We acquired high-precision light curves for four transits of WASP-10~b in 2010. Assuming various limb-darkening laws, we generated best-fit models and redetermined parameters of the system. The prayer-bead method and Monte Carlo simulations were used to derive error estimates.}
   {Three transit light curves exhibit signatures of the occultations of dark spots by the planet during its passage across the stellar disk. The influence of stellar activity on transit depth is taken into account while determining system parameters. The radius of WASP-10~b is found to be no greater than $1.03^{+0.07}_{-0.03}$ Jupiter radii, a value significantly smaller than most previous studies indicate. We calculate interior structure models of the planet, assuming a two-layer structure 
with one homogeneous envelope atop a rock core. The high value of the WASP-10~b's mean density allows one to consider the planet's internal structure including 270 to 450 Earth masses of heavy elements. Our new mid-transit times confirm that transit timing cannot be explained by a constant period if all literature data points are considered. They are consistent with the ephemeris assuming a periodic variation of transit timing. We show that possible starspot features affecting the transit's ingress or egress cannot reproduce variations in transit timing at the observed amplitude.}
   {}


   \maketitle
%
\section{Introduction}

Exoplanets, whose orbits are aligned in such a way that they transit their host stars, constitute a group of worlds whose study can in turn help us to understand the exoplanetary internal structure and atmospheric properties (see e.g. Charbonneau et al. \cite{Charbonneau07}). The transiting exoplanet WASP-10~b (Christian et al. \cite{Christian09}) was discovered by the SuperWASP survey (Pollacco et al. \cite{Pollacco06}). Its mass was initially measured to be $2.96^{+0.22}_{-0.17}$ $M_{\rm{J}}$ (Christian et al. \cite{Christian09}) and then refined to $3.15^{+0.13}_{-0.11}$ $M_{\rm{J}}$ by Johnson et al. (\cite{Johnson09,Johnson10}). Christian et al. (\cite{Christian09}) reported the planetary radius of $1.28^{+0.08}_{-0.09}$ $R_{\rm{J}}$ and note that it is larger than interior models of irradiated giant planets predict. On the basis of a high-precision light curve obtained with the University of Hawaii 2.2-m telescope, Johnson et al. (\cite{Johnson09}) derived a noticeably smaller radius of $1.08 \pm 0.02$ $R_{\rm{J}}$. Further investigations by Krejcov\'a et al. (\cite{Krejcova10}) and Dittmann et al. (\cite{Dittmann10b}) found a planetary radius close to the value reported by Christian et al. (\cite{Christian09}). 

The host star is a faint ($V=12.7$ mag) K5 dwarf located $90 \pm 20$ pc from the Sun in the constellation of Pegasus. It has the effective temperature of $4675 \pm 100$ K (Christian et al. \cite{Christian09}) and the mass of $M_{*}=0.75^{+0.04}_{-0.03}\:M_{\odot}$ (Johnson et al. \cite{Johnson09}). A rotational period of $11.91\pm0.05$ d was determined from photometry affected by starspot-induced brightness modulation (Smith et al. \cite{Smith09}). The similar periodicity of $11.84$ d was detected in radial velocities by Maciejewski et al. (\cite{Maciejewski11}), who re-analysed available spectroscopic data. These authors also found that the orbital eccentricity of the planet is statistically indistinguishable from zero after taking into account the starspot effect. The system was found to be relatively young with an age of $270\pm80$ Myr determined with the gyrochronology method (Maciejewski et al. \cite{Maciejewski11}). 

High-precision transit light curves of a planet moving across a spotted stellar disk offer an outstanding opportunity to map the starspot distribution (Schneider \cite{Schneider}; Silva \cite{Silva}). When a planet occults a starspot area on a star's surface, the flux increases and a characteristic feature in a light curve -- a bump -- is observed, as for example for HD~189733~b (Pont et al. \cite{Pont}), HD~209458~b (Silva-Valio~\cite{SilvaValio}), GJ~1214~b (Berta et al. \cite{Berta}; Carter et al. \cite{Carter}; Kundurthy et al. \cite{Kundurthy}), TrES-1 (Dittmann et al.~\cite{Dittmann10a}), CoRoT-2~b (Wolter et al.~\cite{Wolter}; Huber et al.~\cite{Huber}), or WASP-4~b (Sanchis-Ojeda et al.~\cite{Sanchis}). Unocculted dark spots outside the planetary path across the stellar disk may still be present on the stars's surface and affect the transit light curve. Compared to a spot-free case, spots reduce the effective stellar-disk area, so a stellar radius is underestimated. The observed transits are deeper than a spot-free scenario predicts. This, in turn, results in the overestimate of the planet-to-star radius ratio (Carter et al. \cite{Carter}). The opposite effect may be observed if faculae are considered. These active regions should introduce rather random variations not only in the transit depth, but also in the transit timing and duration.

Interestingly, Maciejewski at al. (\cite{Maciejewski11}) note that transit timing of WASP-10~b cannot be explained by a constant period but by a periodic variation. The existence of a second planet that perturbs the orbital motion of WASP-10~b was postulated as an explanation of the observed transit-time-variation (TTV) signal. A solution with a perturbing planet of mass $\sim$0.1 $M_{\rm{J}}$ and orbital period of $\sim$5.23 d, close to the 5:3 period commensurability, was found to be the most likely configuration. 

In this work, we present the analysis of four new high-quality light curves of WASP-10~b's transits. The aim of this work is to redetermine the system's parameters and verify the existence of the observed TTV signal. We also report on discovering starspot features in transit light curves and show the results of modelling the internal structure of the WASP-10~b planet. 

\section{Observations and data reduction}

\begin{table*}
\caption{The summary of observing runs.} 
\label{table:1}      
\centering                  
\begin{tabular}{c l l c c c c c c}      
\hline\hline                
Run & Date UT     & Time UT        & $N_{\rm{exp}}$ & $X$ & $T_{\rm{exp}}$ (s) & $T_{\rm{cad}}$ (s) & FoV & rms$_{\rm{oot}}$ (mmag)\\ 
\hline                        
   1 & 2010 August 03 & 21:44 -- 01:48 & 165 (175) & $1.94\rightarrow1.03$                & 40 & 67 & $7\farcm83 \times 4\farcm88$ & 1.32 \\ 
   2 & 2010 August 06 & 23:28 -- 03:40 & 213 (217) & $1.24\rightarrow1.01\rightarrow1.04$ & 40 & 66 & $7\farcm83 \times 4\farcm88$ & 1.16 \\
   3 & 2010 September 06 & 22:20 -- 02:04 & 164 (189) & $1.11\rightarrow1.01\rightarrow1.08$ & 50 & 71 & $5\farcm57 \times 3\farcm34$ & 1.06 \\
   4 & 2010 September 09 & 00:12 -- 04:26 & 184 (185) & $1.01\rightarrow1.64$                & 60 & 83 & $6\farcm02 \times 3\farcm73$ & 0.79 \\
\hline                                   
\end{tabular}
\tablefoot{$N_{\rm{exp}}$ -- the number of useful scientific exposures, a total number of exposures in parentheses, $X$ -- airmass changes during the run, $T_{\rm{exp}}$ -- exposure times, $T_{\rm{cad}}$ -- the cadence, FoV -- the field of view, dates in UT at the beginning of the night, rms$_{\rm{oot}}$ -- the data scatter during the out-of-transit phase per point.}
\end{table*}

The $R$-band photometric monitoring of WASP-10~b's transits was performed with the 2.2-m telescope at the Calar Alto Observatory (Spain) during four observing runs in 2010 on August 3 and 6 and September 6 and 9 (programme H10-2.2-011). An additional four hours of monitoring during the out-of-transit phase were run on 2010 September 10 to check a photometric stability. During another two nights, which were granted to the project in 2010 November, poor weather conditions did not allow us to gather scientific data. The Calar Alto Faint Object Spectrograph (CAFOS) in imaging mode was used as a detector. It was equipped with the SITe CCD matrix ($2048 \times 2048$, 24$\mu$m pixel, scale of $0.53$ arcsec per pixel). The telescope was significantly defocused and stellar profiles exhibited a doughnut-like shape. About $5 \times 10^4$ counts per second were recorded for the target, which allowed us to collect up to $3 \times 10^6$ counts in a single exposure. Applying defocusing was expected to minimise random and flat-fielding errors (e.g. Southworth et al. \cite{SouthworthI}). A subframe limiting the effective field of view to the target and a nearby comparison star GSC 02752-00151 and $2 \times 2$ binning mode were used to shorten a readout overhead time to $\sim$25 s. Including overheads, the observations were carried out with a cadence of between 66 and 83 s. Exposure times in individual runs were kept fixed to avoid affecting transit timing. The colour index and brightness of the comparison star, located $3\farcm6$ to the east of WASP-10, were not far from those of the host star that minimised the photometric trends caused by the differential atmospheric extinction. The photometric stability of GSC 02752-00151 was verified by analysing its light curve provided by the SuperWASP survey. We also checked its brightness against fainter stars available in the field and detected no variability. The stellar images were kept exactly at the same position in the CCD matrix during each run thanks to auto guiding. Precise timing was assured by synchronising the computer's clock to Coordinated Universal Time (UTC) using Network Time Protocol software, which is more accurate than 0.1 s. The details of observations are presented in Table~\ref{table:1}.

Weather conditions during run 1 on 2010 August 3 were mainly photometric with occasionally thin clouds in the first half of the observing window. The second half of the run was affected by the light of the rising Moon. A portion of data was lost at the end of an egress phase because of technical problems with dome-slit segments. On 2010 August 6 (run 2), observations were performed in photometric conditions in dark time. The run was again interrupted by the same technical problems during the end of an ingress and the beginning of a flat-bottom phase. The subsequent two runs on 2010 September 6 and 9 were executed in photometric conditions with no moonlight contamination. On September 6, some exposures were lost at the end of the transit owing to a strip of clouds passing across the field. Exposures that were found to be affected by clouds were rejected. 

Our CCD frames were processed using a standard procedure including debiasing but not flat-fielding, which does not improve measurements in the case of defocused and auto-guided observations and may also degrade the data quality  (see e.g. Southworth et al. \cite{Southworth10}). Our tests indeed showed that dividing by a flat-field frame had no detectable effect on a final light curve. The stellar flux was spread over a ring of at least 20 binned pixels ($21\farcs2$) in a diameter. The magnitudes of WASP-10 and the comparison star were determined with differential aperture photometry. After analysing a wide range of apertures, we found that beginning from the aperture radius of 9 binned pixels, the photometric scatter exhibited a flat minimum (or rather a plateau), which is in practice insensitive to greater aperture sizes. The per-exposure photometric uncertainties in the individual measurements were in a range between 0.59 and 0.82 mmag. A visual inspection of the Digitized Sky Survey (DSS) images revealed at least two faint neighbour stars located $12\farcs5$ and $15\farcs9$ away from WASP-10. Their photographic $F$-band magnitudes are 19.6 and 19.9 mag (Zacharias et al. \cite{Zacharias}), hence negligible relative to those of WASP-10, whose brightness in the same band is 11.8 mag. To minimise the neighbour-star contamination, the light curves with minimal apertures for which the lowest scatter was achieved, i.e. 9 binned pixels ($10\farcs6$), were taken as the final ones. The light curves exhibited slight baseline gradients caused mainly by the differential atmospheric extinction. They were approximated by the first or second order polynomials fitted to out-of-transit measurements and then subtracted from individual light curves. This method also removed a possible brightness gradient caused by the host-star variability and the possible effects of moonlight. However, taking into account the small amplitude of the stellar variability (up to $\sim$20 mmag, Maciejewski et al. \cite{Maciejewski11}, see also Section 3.2) and its relatively long timescale of $\sim$12 d (Smith et al. \cite{Smith09}), its contribution remains negligible on the timescale of the observing windows. To estimate the quality of the light curves, the root-mean-square (rms) of data points at the out-of-transit phase was calculated for individual light curves (Table~\ref{table:1}).

\section{Results}

\begin{figure*}
  \centering
  \includegraphics[width=16cm]{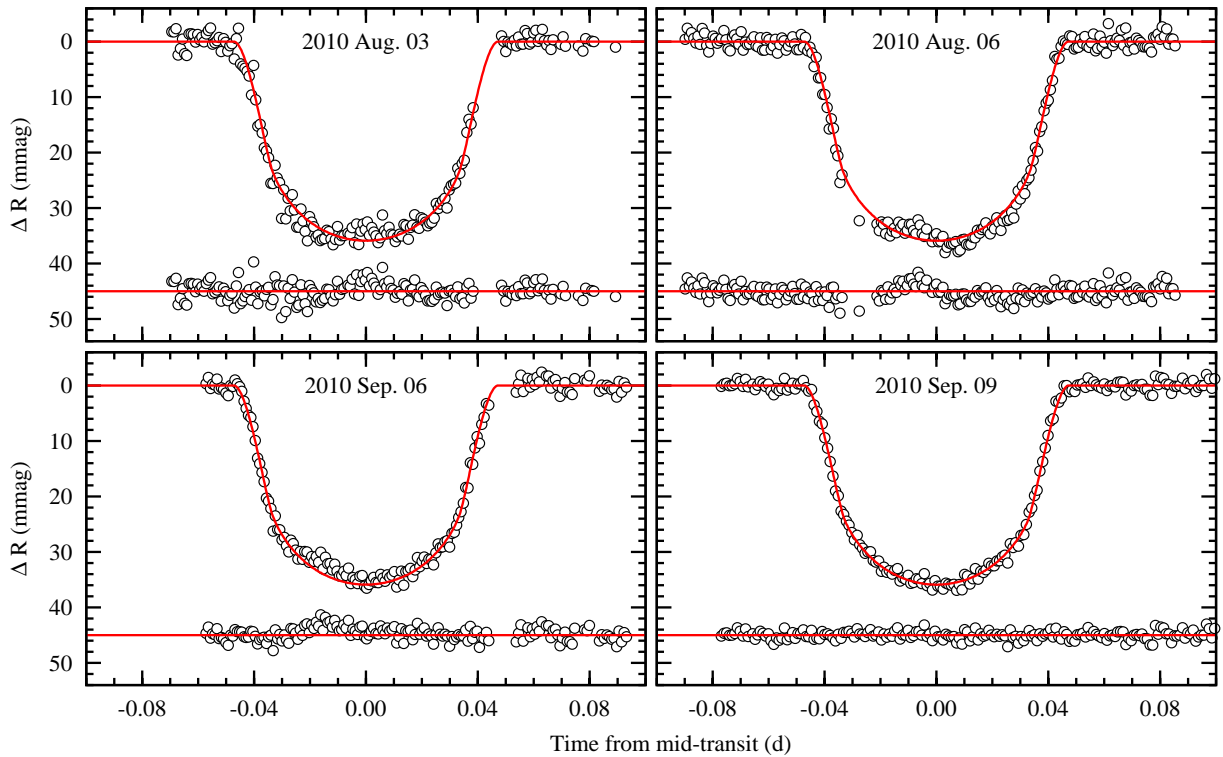}
  \caption{Light curves for four transits of WASP-10~b with models based on the solution obtained for run 4 data and employing the quadratic limb darkening law. The residuals are shown in bottom plots.}
  \label{fig:1}
\end{figure*}

\subsection{Light-curve modelling}

A visual inspection of the first three transit light curves (Fig.~\ref{fig:1}) reveals features that affect the shape of transits and could be attributable to starspot features. The data from the fourth run were found to be of the highest quality exhibiting the lowest data scatter at the out-of-transit phase. This was achieved thanks to photometric conditions during the run and long exposure time. Moreover, these data were found to be almost free of any deformations, thus were used to perform the initial modelling. We used the JKTEBOP code (Southworth et al. \cite{Southwortha, Southworthb}), which is based on the EBOP programme (Eclipsing Binary Orbit Program; Etzel \cite{Etzel}; Popper \& Etzel \cite{Popper}). The initial parameters of the system were taken from Johnson et al. (\cite{Johnson09}). The best-fit model was found by applying the Levenberg-Marquardt least squares procedure. To derive the parameter errors, we used procedures implemented in the JKTEBOP code and described in detail by Southworth (\cite{SouthworthIh}). Firstly, we ran 10000 Monte Carlo (MC) simulations to account for Poisson noise with a bootstrap approach. Secondly, we employed the residual-shift (prayer-bead) method (e.g. D\'esert et al. \cite{Desert11}; Winn et al. \cite{Winn}) to check whether a light curve is affected by correlated noise. A spread range in a given parameter of within 68.3\% was taken as its $1\sigma$ error estimate. 

Six parameters defining the transit shape were fitted directly: the sum of fractional radii of the host star and planet $r_{*}+r_{\rm{b}}$, the ratio of these quantities $k=r_{\rm{b}}/r_{*}$, the orbital inclination $i$, the mid-transit time $T_0$, and the linear $u$ and non-linear $v$ limb-darkening coefficients (LDCs). The fractional radii are defined as $r_{*}=\frac{R_{*}}{a}$ and $r_{\rm{b}}=\frac{R_{\rm{b}}}{a}$, where $R_{*}$ and $R_{\rm{b}}$ are the absolute radii of the star and the planet, respectively, and $a$ is the orbital semi-major axis. A value  $a=0.03781^{+0.00067}_{-0.00047}$ AU (Johnson et al.~\cite{Johnson09,Johnson10}) was used in subsequent calculations. Combining $i$ and $R_{*}$ allowed calculation of the transit parameter $b=\frac{a}{R_{*}}\cos{i}$. The mid-transit times were transformed from Julian Date (JD) based on UTC into Barycentric Julian Date (BJD) based on Barycentric Dynamical Time (TDB) using the on-line converter\footnote{http://astroutils.astronomy.ohio-state.edu/time/utc2bjd.html} by Eastman et al. (\cite{Eastman}). 

We checked linear, quadratic, logarithmic, and square-root limb-darkening (LD) laws and various configurations of LDCs, for which theoretical values were bilinearly interpolated from tables by Van Hamme (\cite{VanHamme}). We considered LDCs kept fixed during fitting, $u$ as a free parameter while $v$ kept fixed, and allowed both LDCs to vary. The best-fit model could be obtained using the square-root LD law with both LDCs being fitted, but the result turned out to be unphysical (total limb darkening at the limb of the star was greater than 1). Keeping $v$ fixed at a theoretical value for the square-root LD law resulted in a poorer goodness of the fit than applying the quadratic LD law for which both LDCs were allowed to vary. Therefore, we used the latter configuration to reproduce the light curve. The MC errors were found to be up to 50\% smaller than those estimated by the prayer-bead method. This indicates that the light curve is affected not only by Poisson noise but also by additional correlated noise. Therefore, our prayer-bead error estimates were taken as our final errors. The sub-millimagnitude precision was achieved with the rms scatter of 0.74 mmag per point. We note that derived LDCs have reasonable values that are consistent to within 1$\sigma$ with theoretical values interpolated e.g. from Claret (\cite{Claret00}) tables for $R_{\rm{C}}$-band filter and a K-type dwarf. Fitting LDCs allows us to account for their contribution to the error budget. Southworth (\cite{SouthworthIh}) shows that keeping LDCs at theoretical values can underestimate the errors of other parameters. 

\begin{figure}
  \centering
  \includegraphics[width=9cm]{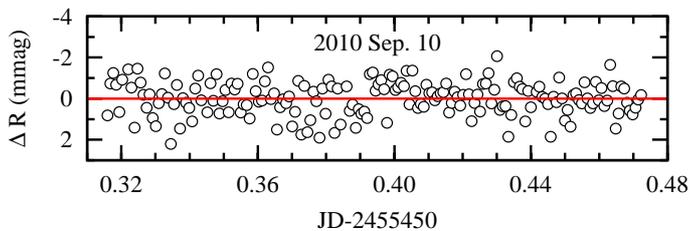}
  \caption{A light curve acquired for WASP-10 during an additional run without transit. The data were processed in a similar way as for transit light curves. The 55-s exposure times were used. }
  \label{fig:1b}
\end{figure}

The best-fit model based on data from the fourth observing run was used as a template and fitted to remaining light curves for which only the mid-transit time was allowed to vary. The residuals are plotted in Fig.~\ref{fig:1}. Both transits observed in August exhibit flux bumps that may be the results of occulting dark-spot areas on the stellar surface by the planetary disk. The negative residuals in run 2 data suggest that in this case the transit's depth is underestimated by the run 4 model. The opposite situation is observed in the light curve acquired on September 6 where, besides the starspot signature, the transit depth tends to be smaller than the run 4 model predicts. Figure~\ref{fig:1b} shows the light curve obtained for WASP-10 on September 10 during an out-of-transit phase. During this almost four-hour-long run, the rms of 0.81 mmag was achieved. The residuals  show no features that could be attributed to starspots. As this run was also affected by clouds, the relatively far more satisfactory photometric stability strengthens the reliability of starspot detection in runs 1--3. We also exclude observing conditions or technical problems as the cause of the observed bumps.

\begin{table*}
\caption{Parameters of transit light-curve modelling.} 
\label{table:2}      
\centering                  
\begin{tabular}{l c c c c}      
\hline\hline                
Parameter & run 1 (Aug. 03) & run 2 (Aug. 06) & run 3 (Sep. 06) & run 4 (Sep. 09)\\ 
\hline                        
Sum of stellar and planetary fractional radii, $r_{*}+r_{\rm{b}}$ & $0.0966^{+0.0030}_{-0.0019}$ & $0.0965^{+0.0022}_{-0.0013}$ & $0.0955^{+0.0022}_{-0.0011}$ & $0.0973^{+0.0023}_{-0.0013}$ \\ 
Planet-to-star radius ratio, $k$ & $0.1596^{+0.0030}_{-0.0029}$ & $0.1607^{+0.0024}_{-0.0032}$ & $0.1565^{+0.0033}_{-0.0022}$ & $0.1586^{+0.0028}_{-0.0027}$ \\ 
Stellar fractional radius, $r_{*}$ & $0.0833^{+0.0022}_{-0.0017}$ & $0.0831^{+0.0016}_{-0.0013}$ & $0.0826^{+0.0017}_{-0.0010}$ & $0.0840^{+0.0019}_{-0.0011}$ \\ 
Planetary fractional radius, $r_b$ & $0.0133^{+0.0007}_{-0.0003}$ & $0.0134^{+0.0006}_{-0.0003}$ & $0.0129^{+0.0006}_{-0.0001}$ & $0.0133^{+0.0004}_{-0.0003}$\\ 
Orbital inclination, $i$ (deg) & $89.0^{+0.7}_{-0.7}$ & $89.2^{+0.7}_{-0.6}$ & $89.5^{+0.4}_{-0.9}$ & $88.9^{+0.4}_{-0.4}$ \\ 
Transit parameter, $b$ ($R_{*}$) & $0.19^{+0.14}_{-0.14}$ & $0.16^{+0.14}_{-0.13}$ & $0.10^{+0.08}_{-0.18}$ & $0.22^{+0.09}_{-0.08}$ \\ 
Transit duration\tablefootmark{a}, $d_{\rm{tr}}$ (min) & $115.9^{+6.2}_{-5.6}$ & $116.3^{+4.9}_{-4.2}$ & $116.5^{+3.2}_{-3.5}$ & $116.1^{+4.9}_{-3.7}$ \\ 
Ingress/egress duration\tablefootmark{b}, $d_{\rm{gr}}$ (min) & $19.2^{+3.8}_{-3.4}$ & $19.2^{+3.5}_{-2.9}$ & $18.4^{+2.3}_{-3.5}$ & $19.4^{+2.3}_{-2.1}$ \\ 
Linear LDC, $u$ & $0.65$\tablefootmark{c} & $0.65$\tablefootmark{c} & $0.65$\tablefootmark{c} & $0.65^{+0.07}_{-0.09}$ \\ 
Non-linear LDC, $v$ & $0.16$\tablefootmark{c} & $0.16$\tablefootmark{c} & $0.16$\tablefootmark{c} & $0.16^{+0.26}_{-0.17}$ \\ 
Mid-transit time, $T_0$ $(\rm{JD_{UTC}})$ & $12.47488^{+0.00010}_{-0.00012}$ & $15.56763^{+0.00015}_{-0.00013}$ & $46.49347^{+0.00007}_{-0.00007}$ & $49.58598^{+0.00008}_{-0.00009}$\\ 
Mid-transit time, $T_0$ $(\rm{BJD_{\rm{TDB}}})$ & $12.47861^{+0.00010}_{-0.00013}$ & $15.57156^{+0.00014}_{-0.00013}$ & $46.49882^{+0.00007}_{-0.00007}$ & $49.59140^{+0.00008}_{-0.00009}$\\ 
Epoch\tablefootmark{d}                   & 341 & 342 & 352 & 353 \\
Timing residual, O--C (d) & $-0.00009$ & $+0.00013$ & $-0.00009$ & $-0.00006$ \\
rms (mmag)            & $1.42$ & $1.11$ & $0.97$ & $0.74$ \\ 
$\chi^2_{\rm{red}}$     & $3.40$ & $1.99$ & $2.06$ & $1.53$ \\ 
\hline                                   
\end{tabular}
\tablefoot{The mid-transit times, $T_0$, are given as 2455400$+$. No spot feature cutting was applied for run 4.}
\tablefoottext{a}{Time between the middle of an ingress and the middle of an egress, defined as $d_{\rm{tr}} = r_{*} P_{\rm{b}} \pi^{-1} \sqrt{1-b^2}$, where $P_{\rm{b}}$ is the orbital period of WASP-10~b taken from Maciejewski at al. (\cite{Maciejewski11}).}
\tablefoottext{b}{Time between the first and second or third and fourth contacts for ingress or egress, respectively, defined as $d_{\rm{gr}} = \frac {r_{\rm{b}} P_{\rm{b}}} {\pi \sqrt{1-b^2}}$, where $P_{\rm{b}}$ is the planet orbital period.}
\tablefoottext{c}{Value adopted from the run 4 fit, permuted by $\pm0.1$ on a flat distribution.}
\tablefoottext{d}{Numbering according to the ephemeris given by Christian et al. (\cite{Christian09}).}
\end{table*}

\begin{figure}
  \centering
  \includegraphics[width=9cm]{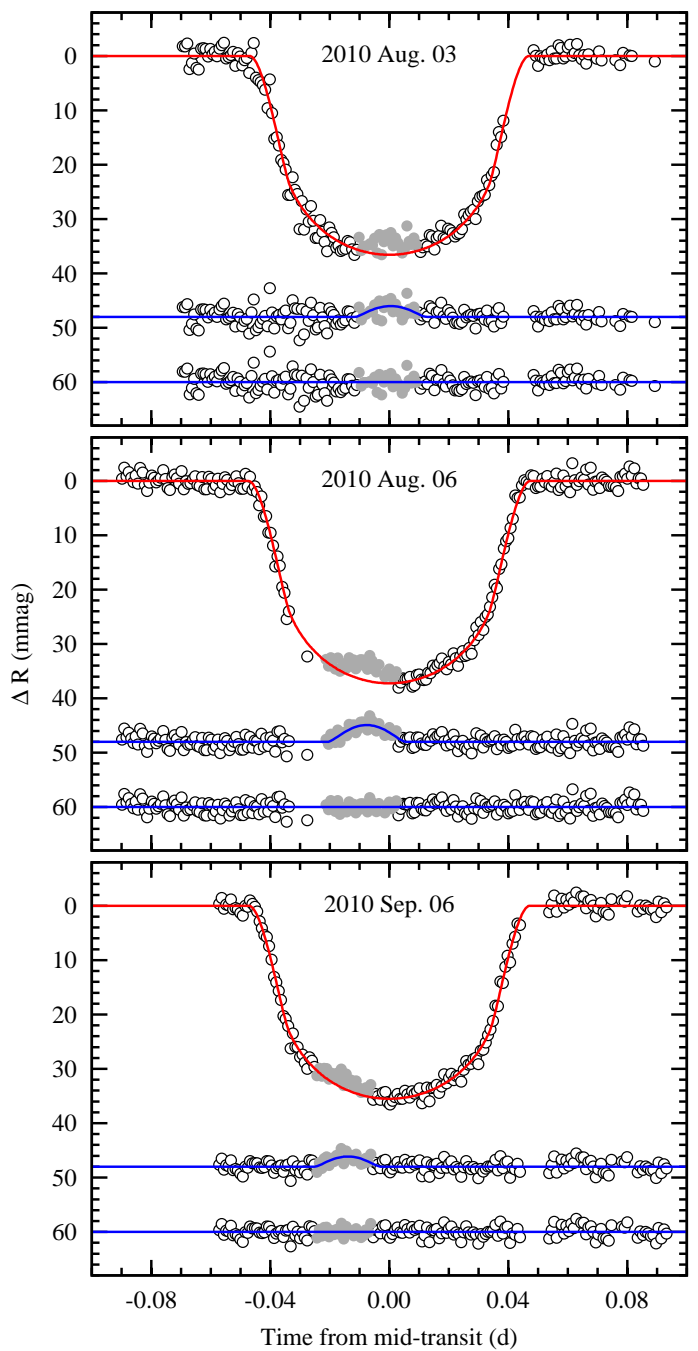}
  \caption{Light curves for three transits of WASP-10~b with individually fitted model light curves after rejecting data points affected by stellar spots (grey points). The residuals that were used to model starspot features are plotted in the middle graphs. In the bottom graphs, the final residuals are plotted.}
  \label{fig:2}
\end{figure}

Variable transit depth may be caused by apparent changes in the effective stellar surface (hence fitted fractional stellar radius), which varies because of the dark-spot coverage. To investigate this scenario, we cut portions of the light curves exhibiting spot signatures and then repeatedly fitted models and estimated the errors. Constraints on the limb darkening were applied because the quality of the reanalysed three light curves is slightly lower than for run 4 data and data sets became significantly incomplete after cutting. We considered only the quadratic LD law and LDCs were kept fixed at values from the run 4 model. Their contribution to the error budget was taken into account by perturbing them by $\pm0.1$ around fixed values with the assumption of a flat distribution. Figure~\ref{fig:2} shows individual best-fit models with residuals, which, if uncut light curves are considered, contain the distilled spot features. Results of the fitting procedure run for individual light curves are collected in Table~\ref{table:2}. 

\subsection{Physical properties of the system}

\begin{figure}
  \centering
  \includegraphics[width=9cm]{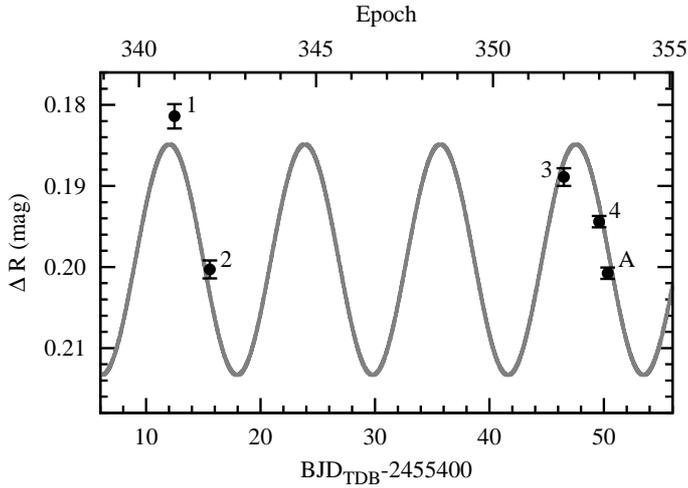}
  \caption{The average brightness of WASP-10 at an out-of-transit phase relative to the comparison star GSC 02752-00151. The continuous line sketches the sinusoidal variation caused by the rotational modulation due to spots. Individual runs are marked with 1-4 numbers, while the additional one during the out-of-transit phase is labelled with A.}
  \label{fig:3}
\end{figure}

The highest-quality light curve of run 4 allowed us to find the best-fit LD law and its LDCs but other system parameters (in particular, the planet-to-star radius ratio $k$) may be affected by stellar activity. Although no spot signatures are visible during this transit, spot complexes may exist on the stellar surface outside the projected path of WASP-10~b. To check this scenario, the average brightness of WASP-10 outside transits was determined in individual nights covered by our observations. Figure~\ref{fig:3} shows the light curve that clearly exhibits data-point scatter with the peak-to-peak amplitude of $\sim$20 mmag. These brightness variations can be closely approximated with a sinusoid whose period is equal to the star rotation period and the peak-to-peak amplitude is $28\pm13$ mmag. Individual magnitude errors (standard deviations) were taken as weights while running the fitting procedure. The derived amplitude is comparable to the greatest value reported so far, i.e. to $20.8\pm1.3$ mmag observed for the 2004 observing season (Maciejewski at al.~\cite{Maciejewski11}). Run 4 was observed during a phase at which WASP-10's brightness was reduced as a result of new starspots appearing on the visible stellar hemisphere. This finding indicates that $k$ derived from these data must be overestimated. The run 1 and 3 data show that even during bright phases, spot features are visible in transit light curves. This finding suggests that the purely photospheric surface was observed in neither run. The photospheric brightness level cannot be determined, so the reliable rescaling of light curves cannot be performed (Czesla et al.~\cite{Czesla}). We note that the LDCs may be affected by stellar activity, e.g. faculae that produce limb brightening and, if numerous, may have a non-negligible contribution to the overall limb darkening.

The presence of starspots located outside the transit chord leads to an overestimate of the transit depth and hence the planet-to-star radius ratio (e.g. Czesla et al.~\cite{Czesla}). Although system parameters are consistent to within $1\sigma$ between individual runs, the smallest value of $k$ was obtained for run 3. The star was then at a bright phase (Fig.~\ref{fig:3}), so a minimal fraction of its surface must have been covered by spots. The run 3 results were chosen as the most representative of the WASP-10 system. They were also expected to be the closest to the real values. Therefore, we used them to redetermine the planetary, stellar, and geometrical properties. Our final results are collected in Table~\ref{table:4}. 

\begin{table*}
\caption{Physical properties of the WASP-10 system derived from light-curve modelling. Literature determinations are cited for comparison.} 
\label{table:4}      
\centering                  
\begin{tabular}{l c c c c}      
\hline\hline                
Parameter & This work & Christian et al. (\cite{Christian09}) & Johnson et al. (\cite{Johnson09,Johnson10}) & Krejcov\'a et al. (\cite{Krejcova10})\\ 
\hline                        
Planet radius, $R_{\rm{b}}$ $(R_{\rm{J}})$       & $1.03^{+0.07}_{-0.03}$ & $1.28^{+0.077}_{-0.091}$ & $1.08\pm0.02$ & $1.22\pm0.05$ \\ 
Planetary mean density, $\rho_{\rm{b}}$ $(\rho_{\rm{J}})$   & $2.94^{+0.46}_{-0.25}$ & $1.43^{+0.31}_{-0.29}$ & $2.35\pm0.15$ & -- \\ 
Planet gravitational acceleration, $\log g_{\rm{b}}$ (cgs) & $3.88^{+0.04}_{-0.01}$ & $3.62\pm0.06$ & $3.828\pm0.012$ & -- \\ 
Equilibrium temperature, $T'_{\rm{eq}}$ (K) & $950^{+30}_{-26}$ & $1119^{+26}_{-28}$ & $1370\pm50$ & -- \\ 
Orbital inclination, $i$ (deg) & $89.5^{+0.4}_{-0.9}$ & $86.9^{+0.6}_{-0.5}$ & $88.49^{+0.22}_{-0.17}$ & $87.3\pm0.1$ \\
Transit parameter, $b$ $(R_{*})$ & $0.10^{+0.08}_{-0.18}$ & $0.568^{+0.054}_{-0.084}$ & $0.299^{+0.020}_{-0.043}$ & -- \\ 
Planet-to-star radius ratio, $R_{\rm{b}}/R_{*}$  & $0.1565^{+0.0033}_{-0.0022}$ & $0.170\pm0.002$ & $0.15918^{+0.00050}_{-0.00115}$ & $0.168\pm0.001$ \\ 
Star radius, $R_{*}$ $(R_{\odot})$ & $0.67^{+0.03}_{-0.02}$ & $0.775^{+0.043}_{-0.040}$ & $0.698\pm0.012$ & $0.75\pm0.03$ \\
Scaled semi-major axis, $a/R_{*}$ & $12.11^{+0.24}_{-0.15}$  & $10.23^{+0.90}_{-0.92}$  & $11.65^{+0.09}_{-0.13}$ & $10.64\pm0.12$ \\ 
Mean star density, $\rho_{*}$ $(\rho_{\odot})$ & $2.48^{+0.26}_{-0.17}$ & $1.51^{+0.25}_{-0.20}$ & $2.20\pm0.063$ & -- \\ 
Star gravitational acceleration, $\log g_{*}$ (cgs) & $4.66^{+0.06}_{-0.04}$ & $4.51^{+0.06}_{-0.05}$  & $4.627^{+0.0101}_{-0.0093}$ & -- \\ 
\hline                                   
\end{tabular}
\tablefoot{The equilibrium temperature, $T'_{\rm{eq}}$, was derived by assuming the effective temperature of the host star $T_{\rm{eff}}=4675\pm100$ K (Christian et al. \cite{Christian09}) and simplified relation $T'_{\rm{eq}} = T_{\rm{eff}} \sqrt{r_{*}/2}$ (Southworth \cite{SouthworthIIIh}). Some quantities that were needed for calculations and could not be determined from the light curve analysis (e.g. planetary mass, semi-major axis) were taken from Johnson et al. (\cite{Johnson09,Johnson10}).}
\end{table*}

Our parameters of the WASP-10 system differ by $0.6$--$1.9\sigma$ from those reported by Johnson et al. (\cite{Johnson09,Johnson10}) with only the planet gravitational acceleration value differing by more than $2\sigma$.  The only exception is $T'_{\rm{eq}}$ whose value by Johnson et al. (\cite{Johnson09,Johnson10}) is $\sim$$\sqrt{2}$ times greater than our determination. The reason for this discrepancy is the factor of $\sqrt{2}$ missing in the formula defining the equilibrium temperature in Johnson et al. (\cite{Johnson09}). The high-quality light curve analysed by Johnson et al. (\cite{Johnson09}) exhibits no signs of spot occultations and, as an extrapolation of the brightness modulation in Fig.~\ref{fig:3} suggests, was acquired during a bright phase, i.e. when the spot coverage was at its smallest. 
System parameters reported by Christian et al. (\cite{Christian09}) and Krejcov\'a et al. (\cite{Krejcova10}) differ by $1.3$--$5.4\sigma$ from those derived by us.  
The planet-to-star radius ratio and star radius have the greatest impact on other parameters. The light curve analysed by Dittmann et al. (\cite{Dittmann10b}), whose $k=0.16754\pm0.00060$ was found to be close to values of Christian et al. (\cite{Christian09}) and Krejcov\'a et al. (\cite{Krejcova10}), exhibits some features attributable to starspots. Studies by Christian et al. (\cite{Christian09}) and Krejcov\'a et al. (\cite{Krejcova10}) based on average light curves that blurred the influence of activity-induced features. Although our radius estimate of $R_{\rm{b}}=1.03^{+0.07}_{-0.03}$ $R_{\rm{J}}$ is the smallest determined so far, it is de facto the maximal limit because we could take into account only a fraction of the whole spot contribution. If the greatest value of $k$ were considered (i.e. for the run 2 data set), the planetary radius would be greater by 3\%, i.e. well within error bars. Hence, the difference seems to be negligible in practise. Using the Roche model, Budaj (\cite{Budaj}) has found that the shape of WASP-10~b is not far from the sphere. The ratio of the planetary radii at the sub-stellar point to that at the rotation pole is 1.00132. Thus, the planet radius determined from the transits is practically the same as the effective radius which is defined as the radius of the sphere with the same volume as the Roche surface used in theoretical calculations.
Hartman (\cite{Hartman10}) discovered a correlation between the surface gravitational acceleration of hot Jupiters and the activity of host stars. According to this relation, the surface gravity of WASP-10~b, $\log g_{\rm{b}}=3.88^{+0.04}_{-0.01}$ (cgs), predicts a strong chromospheric activity of the WASP-10 star with $\log R'_{\rm{HK}}>-4.3$. The star is expected to have pronounced emission features in the H
and K Ca II lines that make it an interesting target for spectroscopic studies.

The satisfactory consistency is obtained when the derived gravitational acceleration of the host star, $\log g_{*} = 4.64^{+0.06}_{-0.04}$, is compared to theoretical predictions. We interpolated the solar-metallicity isochrones by Girardi et al. (\cite{Girardi}) to the age of the WASP-10 system and derived a value of $\sim$4.69, which is within 1$\sigma$ error bars of the value reported in this work. The value derived by Johnson et al. (\cite{Johnson09,Johnson10}) is close to the theoretical one, while the result brought by Christian et al. (\cite{Christian09}) is significantly underestimated.

\subsection{Internal structure}
We calculate interior structure models of WASP-10~b assuming a two-layer structure 
with one homogeneous envelope atop a rock core. We aim to determine the resulting
possible mass of heavy elements in both the core and the envelope
as a function of the planet radius, which we  vary between the small value of
1.03 $R_{\rm J}$ (this work) and 1.28 $R_{\rm J}$ found by Christian et al. (\cite{Christian09}),
while the mass is kept invariant at 2.96 $M_{\rm J}$ (Christian et al.~\cite{Christian09}).

For H and He we use the interpolated equation of state (EOS) by Saumon et al. (\cite{Saumon}), for heavy 
elements in the envelope we scale the He EOS by a factor of four in density,
and for rocks in the core we use the rock EOS by Hubbard \& Marley (\cite{Hubbard}).
This is a choice of convenience, as we are interested in the response of
the planet's metallicity to the planet's radius more than to achieve 
a high accuracy in these numbers for a single model. The
composition of heavy elements is essentially unknown, and a different choice
of the EOS of metals can affect the calculated mass of metals by $\sim 30\%$.

An interior model of a close-in exoplanet cannot stand alone without a model for the irradiated model atmosphere, as the atmosphere determines 
the entropy of a gas giant's interior. The hotter the atmosphere, the larger the entropy, 
and the more heavy elements are required to meet the observed transit radius. Common features of the atmospheric pressure--temperature
profile of evolved hot Jupiters around Sun-like stars are the presence of
a temperature inversion at high altitudes at sub-mbar pressures, the sub-adiabatic temperature profile in the radiative region where the
transit radius is observed (Fortney et al.~\cite{Fortney2003}), the development of an isothermal region around 1-100 bars depending on an age and orbital distance, 
and finally the onset of the adiabatic, convective interior, 
which extends down to the core in the likely absence of internal layer boundaries 
in hot exoplanets. We use Fig.~3 in Fortney et al. (\cite{Fortney}) to construct an
atmosphere profile by interpolation, where we increase the orbital distance
from $a=0.037$ AU 
to 0.07 AU in order to conserve the total energy flux received by
WASP-10~b from its parent K5 star. The resulting isothermal region of
our atmosphere profile is at 1400 K and begins at 2 bars. 
As the isothermal region extends deeper into the planet with increasing
planet age (Fortney et al.~\cite{Fortney}), we expect a shallow or even absent 
isothermal region for the young WASP-10~b. We use Fig.~2 in Fortney et al. (\cite{Fortney}) 
to estimate the onset of the adiabatic interior, which we parametrise by a 
pressure level $P_{\rm ad}$. The equilibrium temperature of WASP-10~b is 850-980 K
for albedos between 0.3 and zero. Hence, an effective temperature of
1000 K is certainly a lower limit for WASP-10~b. For our choice of planet radii,
the surface gravity is $\log g_{\rm{b}} = 3.62$ (i.e. $g_{\rm{b}} = 42$ m~s$^{-2}$ from Christian et al.~\cite{Christian09}) and $\log g_{\rm{b}} = 3.88$ (i.e. $g_{\rm{b}} = 76$ m~s$^{-2}$ determined by us). For a young hot Jupiter  with $\log g = 3.6$ ($g=40$ m~s$^{-2}$) and $T_{\rm eff}=1000$ K at 0.1 AU, Fortney et al. (\cite{Fortney}) predict 
$P_{\rm ad}\approx 3$ bar. We vary $P_{\rm{ad}}$ from 0.5 to 10 bars to cover the
uncertainty related to a possible higher $T_{\rm eff}$ and larger gravity.
The procedure for calculating the internal profile and the core mass is the same 
as described in Nettelmann et al. (\cite{Nettelmann}). The envelope heavy element mass fraction
is set to the solar value of 1.5\%, which also agrees with the stellar metallicity
to within the error bars. This choice gives the maximum core mass as a higher
envelope metallicity reduces the core mass when all other parameters are
not varied.

\begin{figure}
  \centering
  \includegraphics[width=9cm]{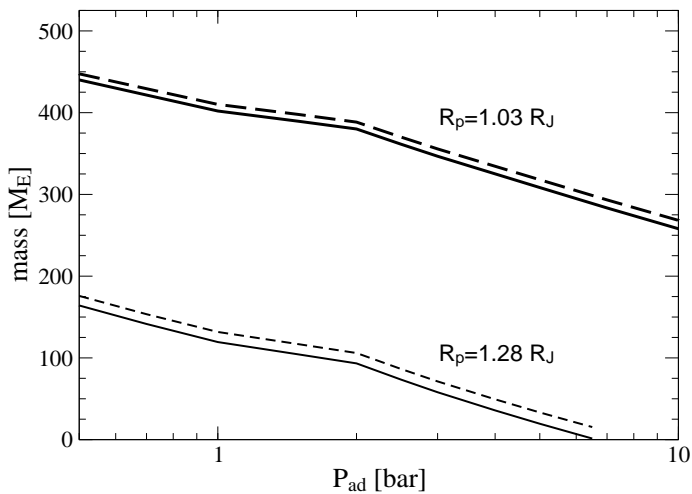}
  \caption{Core mass (solid lines) and total mass of metals (dashed lines) of WASP-10b
interior models with solar metallicity envelope, both in $M_{\oplus}$. The models differ in the onset
of the adiabatic interior as parametrised by the pressure level $P_{\rm ad}$. 
The observed transit radius of this work requires to increases the total mass of metals by 
$\sim$$300\:M_{\oplus}$ compared to the larger radius from the discovery paper.}
  \label{fig:McMz}
\end{figure}

Figure~\ref{fig:McMz} shows the results of our calculations. The total mass of metals $M_{\rm{Z}}$ of 
270--450$\:M_{\oplus}$ is needed to reproduce the radius value reported in this work.
This large amount is challenged by the core-accretion formation where the maximum 
mass of available heavy elements is $80 \:M_{\oplus}$ for the parameters of WASP-10~b
(Leconte et al.~\cite{Leconte}). In contrast, the larger radius scenario
predicts a mass of heavy elements that is smaller by $\sim$$300\:M_{\oplus}$, which corresponds
to a heavy element enrichment of $(12\pm 11)$ times a stellar value, i.e. more in line with Jupiter.

\subsection{Spot feature modelling}

\begin{table*}
\caption{Spot-occultation parameters.} 
\label{table:3}      
\centering                  
\begin{tabular}{l c c c}      
\hline\hline                
Parameter & run 1 (Aug. 03) & run 2 (Aug. 06) & run 3 (Sep. 06) \\
\hline                
Height (mmag) & $1.99\pm0.44$ & $3.07\pm0.32$ & $1.87\pm0.31$ \\
Duration (min)  & $34.3\pm7.9$ & $37.2\pm4.0$ & $31.6\pm5.5$  \\
Mid-occultation time -- $T_0$ (d)  & $+0.0008\pm0.0014$ & $-0.0074\pm0.0007$ & $-0.0141\pm0.0010$  \\
Percentage of stellar disk area\tablefootmark{a} & $0.60\pm0.13$ & $0.9\pm0.1$ & $0.6\pm0.1$ \\
Longitude\tablefootmark{b} ($^\circ$) & $+1.1$ & $-10.5$ & $-20.3$ \\
rms (mmag) & $1.43$ & $1.07$ & $0.95$  \\
\hline                        
\hline                                   
\end{tabular} \\
\tablefoottext{a}{The spot contrast of 0.7 (typical for the Sun) was assumed.}
\tablefoottext{b}{Defined as $-90^\circ$ at ingress, $0^\circ$ at mid-transit time, and $+90^\circ$ at egress.}
\end{table*}

The spot-occultation features (Fig.~\ref{fig:2}) may be approximated with a simplified transit-like model assuming a flux bump instead of a flux drop. To model these features, an approach similar to that presented by Kundurthy et al. (\cite{Kundurthy}) was used. The bumps were modelled with the \textsc{occultsmall} routine of Mandel \& Agol (\cite{mandelalgol02}) and the Levenberg--Marquardt non-linear least squares fitting algorithm provided by the Exoplanet Transit Database\footnote{http://var2.astro.cz/ETD/} (Poddan\'y et al. \cite{poddanyetal10}). In this case, the limb darkening was assumed to be zero and the mid-occultation time, occultation height, and duration counted from the begin to the end of the phenomenon were fitted. The error estimates of these parameters were taken from a covariance matrix. The best-fitting models are plotted in Fig.~\ref{fig:2} (middle plots) and occultation parameters are printed in Table~\ref{table:3}. The final residuals are shown in the bottom plots of Fig.~\ref{fig:2}. The height of the bumps was found to be between 2 and 3 mmag. The durations of the events are slightly shorter than a sum of transit ingress and egress durations equal to $\sim$40 min (see Table~\ref{table:2}). Moreover, no flat top phase is visible. These findings are most consistent with the grazing scenario and non-central occultations of single spots whose sizes are comparable to or larger than the planetary disk projected onto the stellar surface. The occulted starspots were found to be located near the centre of the stellar disk at longitudes within $20^\circ$ of the line of sight. The height of the bumps was determined at the 5--9-$\sigma$ level. The significance of these bumps can be estimated from the ratio of their height to the final light-curve scatter represented by the rms. The detections of starspot signatures in run 1, 2, and 3 data were found at the 1.4-, 2.9-, and 2.0-$\sigma$ level. The run 2 event was the most pronounced while the bump in the run 1 light curve was affected by lower quality data. While reducing data, we excluded instrumental effects being responsible for the observed features in the light curves.

The flux bumps can be translated into the area of a spot occulted by the planetary disk. Assuming that the spot contrast is equal to 0.7, a value typical of the Sun, the occulted spot area was found to be in a range between 0.6\% and 0.9\% of the stellar surface (or 25\% and 37\% of the planetary disk). These inhomogenities alone cannot produce the observed rotational modulation in photometry. To reproduce the observed amplitude of $\sim$$28$ mmag, a distinct spot of a size of either $\sim$9\% or $\sim$2.6\% of the stellar disk would be needed for a solar contrast or completely dark spot, respectively. As the amplitude of the rotational modulation was found to be one of the greatest observed for WASP-10 so far, the radial velocity variations with a semi-amplitude much greater than $\sim$70 m~s$^{-1}$, a value reported by Maciejewski et al. (\cite{Maciejewski11}), would be expected if simultaneous spectroscopic observations were done. The starspot features observed in the transit light curves are consistent with the observed radial velocity scatter and the photometric variability of the host star.

\subsection{Transit timing}

\begin{figure*}
  \centering
  \includegraphics[width=17cm]{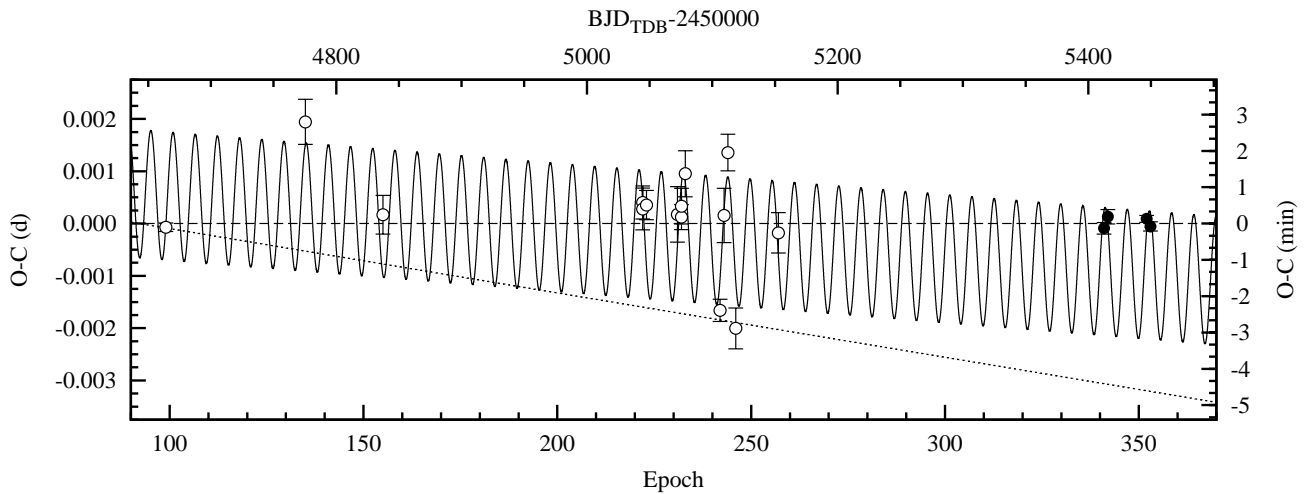}
  \caption{The O--C diagram for mid-transit times of WASP-10~b. The mid-transits published in the literature are marked with open symbols. The filled ones denotes new mid-transit times determined by us. The periodic signal predicted by Maciejewski at al. (\cite{Maciejewski11}) is sketched with a continuous line. The dashed line shows the refined linear ephemeris and the dotted one marks the linear ephemeris provided by Dittmann et al. (\cite{Dittmann10b}).}
  \label{fig:4}
\end{figure*}

Four new mid-transit times in conjunction with the literature data allowed us to refine the linear ephemeris. We obtained the mid-transit time for initial epoch $T_0=24554357.85790\pm0.00010$ $\rm{BJD}_{\rm{TDB}}$ and the orbital period $P_{\rm{b}}=3.09272963\pm0.00000035$ d. Individual timing errors were taken as weights while running the linear regression. Our four new mid-transit times could be used to verify the TTV signal reported by Maciejewski et al. (\cite{Maciejewski11}). Figure~\ref{fig:4} shows the observation minus calculation (O--C) diagram plotted for all data according to the new ephemeris. Although our four new points are located close to the linear ephemeris, they also agree with the ephemeris given by Maciejewski at al. (\cite{Maciejewski11}). The remaining two runs, which were granted to us and then lost because of poor weather conditions, had been expected to cover epochs during opposite phases of the TTV signal and could have been used to determine precisely its amplitude. It is worth noting that all four transits happened $\sim$5 min later than the linear ephemeris by Dittmann et al. (\cite{Dittmann10b}) predicts. This finding confirms the TTV signal of WASP-10~b and is consistent with the model proposed by Maciejewski et al. (\cite{Maciejewski11}).  

To estimate the influence of starspot occultations on transit timing, the run 4 light curve was artificially deformed by injecting a flux bumps randomly located in the time of transit. The duration of artificial occultations was assumed to be $\sim$40 min and its height $3.07$ mmag, i.e. values of the most prominent feature observed in run 2. Two separate scenarios were considered: an affected first half of a flat bottom and ingress/egress. In the first case, 2000 deformed light curves were produced and then fitting procedure was performed. The sum of stellar and planetary fractional radii, planet-to-star radius ratio, orbital inclination, and mid-transit time were allowed to vary while the quadratic LDCs were kept fixed at values from the run 4 model. The mean difference in $T_0$, $\delta T_0$, was found to be negligible with a value of $0.00006\pm0.00004$ d (i.e. $0.09 \pm 0.06$ min), where the error is the sample standard deviation. This finding was also confirmed by our analysis of run 1--3 light curves: the mid-transit times derived from original and starspot-feature-cut light curves were found to be similar well within $1\sigma$ error bars. The planet-to-star radius ratio was found to be significantly smaller by about $5\sigma$. This result meets expectations because injected bumps increase the averaged brightness during the flat bottom phase, hence decrease the averaged transit depth.

The procedure was repeated for ingress phase based on 1000 deformed light curves and yielded $\delta T_0 = 0.00062 \pm 0.00005$ d (i.e. $0.89 \pm 0.07$ min), indicating that the apparent mid-transit times happened noticeably later than the spotless ephemeris predicted. The orbital inclination was found to be underestimated leading to a greater value of the transit parameter. Moreover, transits appeared to be shorter on average by $1.5\pm0.6$ min. Because of the symmetry of the phenomenon, the same results can be reached for the deformed second part of the flat bottom and egress. In these cases, the apparent mid-transit times would happen earlier. This phenomenological approach gives the upper limits to any possible deviations in timing. As Sanchis-Ojeda et al. (\cite{Sanchis}) show, the influence of occulted starspots located close to the stellar limb is smaller by a factor of 3--5 because of the effects of limb darkening and geometrical foreshortening.

\section{Discussion}

Three of our four transit light curves exhibit signatures of starspots, while Maciejewski et al. (\cite{Maciejewski11}) detected no variation in the transit shapes of nine light curves. Most of their data sets have a data scatter of around or larger than 2 mmag. Such a precision seems to be insufficient for detecting flux bumps reported by us. Their only light curve with the data scatter comparable to ours exhibits no features that could be related to an occultation of an active region. However, this finding should be treated with caution as a final part of the egress and out-of-transit phase was not covered by observations.

In the case of defocused images, the light contamination from nearby faint stars may make a transit shallower and hence lower the planet-to-star radius ratio. In the neighbourhood of WASP-10, there are two faint stars that could contribute light to the WASP-10's flux (see Section 2). Assuming that the total flux of these neighbouring objects (based on their photographic magnitudes) contributed to the WASP-10 light curve, the planetary radius would be underestimated by 0.07\% -- a value negligible in practice.     

\begin{figure}
  \centering
  \includegraphics[width=8cm]{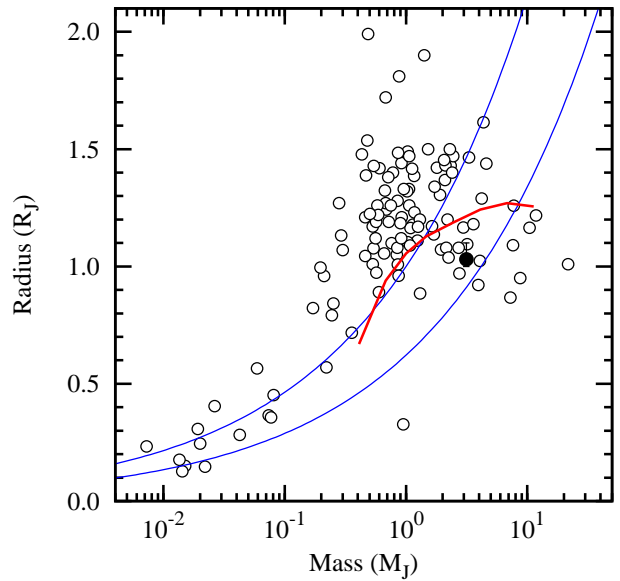}
  \caption{The mass-radius diagram for currently known transiting exoplanets (grey dots). The black spot marks WASP-10~b. The density contours of Jupiter (upper) and Earth (lower) are plotted with blue lines. The red line sketches the planetary isochrone for a 100 $M_{\oplus}$ core planet, the interpolated equivalent semi-major axis of WASP-10~b, and the age of 300 Myr (Fortney et al. \cite{Fortney}).}
  \label{fig:5}
\end{figure}

The high mean density of WASP-10~b ($2.94^{+0.46}_{-0.25}$ $\rho_{\rm{J}}$) in conjunction with the relatively young age of the WASP-10 system testifies for the planet's heavy-element core of a significant mass, as discussed by Johnson et al. (\cite{Johnson09}). The models of internal structure by Fortney et al. (\cite{Fortney}) and Baraffe et al. (\cite{Baraffe08}) predict a substantial core of the mass $>$$100$ $M_{\oplus}$ for WASP-10~b. In the mass-radius plot (Fig.~\ref{fig:5}), WASP-10~b is located in a sparsely populated area covered by deflated exoplanets. Our models of the WASP-10~b's internal structure show that for the small radius reported in this work, the total mass contained in metals  (270--450$\:M_{\oplus}$) is much higher than that for Jupiter (12--47$\:M_{\oplus}$; Saumon \& Guillot~\cite{SaumonGuillot}) and in terms of planet mass fraction ($M_{\rm{Z}}/M_{\rm{planet}}=0.28-0.47$), WASP-10~b is located between Jupiter (0.04--0.12) and HD149026~b (0.43--0.72, Ikoma et al.~\cite{Ikoma}). Since $M_{\rm{Z}}$ is quite insensitive to the assumed distribution of heavy elements between the core and envelope, the $M_{\rm{Z}}$ values can be used to derive an overall enrichment of $35\pm 25$ times the stellar metallicity $Z_{*}$ for which [M/H]=$0.02\pm 0.3$ (Christian et al.~\cite{Christian09}). The extremely high core mass of $\geq 250\:M_{\oplus}$ indicates that the envelope is not of solar metallicity but of a few times $Z_{\odot}$, at least.

If the spins of both the stellar rotation and planetary orbital motion are roughly parallel to each other, this configuration may allow one to either determine or refine a period of stellar rotation by analysing the starspot distribution probed by a transiting planet (Silva-Valio~\cite{SilvaValio}; Silva-Valio \& Lanza \cite{SilvaLanza}; Huber et al.~\cite{Huber}). In a significant fraction of extrasolar systems, both axes were found to be misaligned (see e.g. H\'ebrard et al. \cite{Hebrard10}) and this could also apply to the WASP-10 system. The ratio of periods of WASP-10~b's orbital motion to stellar rotation is close to 1:3.8, hence the occultations of the same spot near the centre of the stellar disk (as in runs 1--3) are impossible to observe during two consecutive transits. However, assuming that the active regions exist longer than a few stellar rotation periods and the system is not significantly tilted, we found that features observed during runs 1 and 3 (the time span of about three rotation periods) could be attributable to the same spot complex. In this case, mid-occultation times would refine the rotation period of WASP-10 to $11.85$ d and $10.87$ d for prograde and retrograde orbits, respectively. The former value agrees with the period known from photometric and spectroscopic observations.

About 20\% of the stellar disk is probed during transits. If the system is aligned, the transit chord covers a latitudinal band at the stellar equator (up to about $\pm10^\circ$ in latitude). Systematic and long-timescale studies of the transit profile morphology would allow one to investigate a spot distribution during the cycle of stellar activity.  

A typical full width at half-maximum for WASP-10 cross-correlation functions (CCFs) obtained with the SOFIE spectrograph in its high-efficiency (HE) observing mode by cross-correlating the WASP-10 spectra with a K5V template was found to be in a range of 10.2--10.4 km~s$^{-1}$ (Christian et al. \cite{Christian09}). This translates into the mean projected stellar rotational velocity $v \sin i_{*} = 4.1 \pm 1.0$ km~s$^{-1}$. Here we used the calibration of the projected rotational velocity for the HE observing mode of the SOPHIE spectrograph by Boisse et al. (\cite{Boisse}) with a typical error of 1.0 km~s$^{-1}$. We also assumed the star's intrinsic colour index to be $(B-V)_0=1.15\pm0.10$ mag, a value derived from the spectral type and Schmidt-Kaler (\cite{Schmidt}) calibration for main sequence stars. Here the formal error estimate reflects the uncertainty in one spectral class. We also checked the consistency of the WASP-10's spectral type with its near-infrared colour indices from the 2-Micron All Sky Survey (2MASS, Skrutskie et al.~\cite{Skrutskie}) using the relation given by Ducati et al. (\cite{Ducati}). Using the rotation period determined by Smith et al. (\cite{Smith09}), the minimal radius of the star appears to be $0.98^{+0.31}_{-0.30}$ $R_{\odot}$. This value is by 40\% greater than our photometric analysis predicts. Both values are marginally consistent within the error bars. Additional spectroscopic observations are needed to remove this discrepancy and refine $v \sin i_{*}$ for WASP-10. We note that a larger stellar radius induces a greater planet size, which, in turn, could help to reduce the mass of WASP-10~b`s core in the internal structure models.

The durations of the individual transits were found to be the same for all four runs that shows that the ingresses or egresses of light curves are unaffected by occultations of starspots located close to the stellar disk limb. Limiting the O--C diagram to points based on high-quality photometry, it is obvious that the linear fit is an unsatisfactory solution because some data points deviate by up to 8$\sigma$. We re-analysed high-quality light curves available in the literature, i.e. from epochs 99 (Johnson et al. \cite{Johnson10}) and 242 (Maciejewski et al. \cite{Maciejewski11}), to measure the variation in the transit duration $d_{\rm{tr}}$. Following the rule of selecting the largest errors, the MC errors were taken as the final ones for the epoch 99 light curve. In the second data set, a linear trend, clearly visible during the out-of-transit phase, was fitted simultaneously. Here the prayer-bead errors were found to be greater and taken in further calculations. We found $d_{\rm{tr}}$ equal to $115.9^{+5.1}_{-4.2}$ and $116.0^{+8.9}_{-7.7}$ min for epoch 99 and 242 data sets, respectively. Both values are close to our determinations, a result that allows us to exclude an ingress/egress deformation by an occulting starspot as a source of the TTV signal. We also note that the redetermined mid-transit times were found to be consistent with the literature values and to lie well within the 1$\sigma$ error bars.

\section{Conclusions}

The signatures of WASP-10's activity have been observed in our high-precision transit light curves. Our analysis of system parameters, which takes stellar activity into account, has indicated that the radius of WASP-10~b is smaller than most prior studies report. This planet is three times more massive than Jupiter but has a similar radius. Its heavy-element mass, which is between 270 and 450$\:M_{\oplus}$, needs to be considered in the internal structure models to reproduce the planetary radius. Our new data points confirm that transit timing cannot be explained by a constant period if all literature data points are considered. They are consistent with the ephemeris assuming a periodic variation in the transit timing. We have found no means of explaining the observed TTV signal with light curve deformations caused by starspots on the stellar surface. This finding supports a scenario in which the second planet perturbs the orbital motion of WASP-10~b. However, we emphasis that additional high-quality transit light curves will be required to help us establish the true nature of this intriguing system.

\begin{acknowledgements}
We are grateful to the anonymous referee for valuable comments and suggestions, which enhanced our research. We thank the editor, Tristan Guillot, for his remarks, which helped to improve the manuscript. We also thank staff of the Calar Alto Astronomical Observatory for their support during observing runs. GM and GN acknowledge the financial support of the Polish Ministry of Science and Higher Education through the Iuventus Plus grant IP2010 023070. GM also acknowledges support from the EU in the FP6 MC ToK project MTKD-CT-2006-042514. SR and RN would like to thank the German national science foundation Deutsche Forschungsgemeinschaft (DFG) for support through the grant NE 515/33-1 in the DFG Special Priority Program SPP 1385 on \textit{The First 10 Million Years of the Solar System - a Planetary Materials Approach}. MS and RN would like to thank DFG for support in project NE 515/36-1. NN acknowledges discussions with J.J.Fortney on planetary atmospheres. CA and RN would like to thank DFG for support in projects NE 515/30-1 and 35-1, the latter as part of DFG SPP 1385. We would also like to thanks our Co-PIs
in projects DFG NE 515/33-1, 35-1, and 36-1: W.Duschl, P.Hauschildt, A.V.Krivov, and A.Hatzes. GM, SR, and RN acknowledge support from the DAAD PPP--MNiSW project 50724260--2010/2011 \textit{Eclipsing binaries in young clusters and planet transit time variations}. Finally, we would like to thank the DFG for financial support for the observing runs in programmes NE 515/40-1 and 41-1. 

We have used data from the WASP public archive in this research. The WASP consortium comprises of the University of Cambridge, Keele University, University of Leicester, The Open University, The Queen's University Belfast, St. Andrews University and the Isaac Newton Group. Funding for WASP comes from the consortium universities and from the UK's Science and Technology Facilities Council.

This publication makes use of data products from the Two Micron All Sky Survey, which is a joint project of the University of Massachusetts and the Infrared Processing and Analysis Center/California Institute of Technology, funded by the National Aeronautics and Space Administration and the National Science Foundation.

\end{acknowledgements}

\end{document}